# Dynamic Modeling of Precipitation in Electrolyte Systems

Niklas Kemmerling,[a] Sergio Lucia[a]

*[a]TU Dortmund, Emil-Figge Str. 70, Dortmund 44227, Germany*
*niklas.kemmerling@tu-dortmund.de*

## Abstract

This study presents a dynamic modeling approach for precipitation in electrolyte systems, focusing on the crystallization of an aromatic amine through continuous processes. A novel model, integrating equilibrium and crystallization kinetics, is formulated and applied to a continuous oscillatory baffled reactor. The approach assumes rapid equilibrium establishment and is formulated as a set of differential algebraic equations. Key features include a population balance equation model to describe the particle size distribution and the modeling of dynamically changing equilibria. The predictions of the dynamic model show good agreement with the available experimental measurements. The model is aimed at aiding the transition from a batch process to continuous process by forming the basis for numerical optimization and advanced control.

**Keywords**: dynamic modeling, reaction model, continuous crystallization, Continuous Oscillatory Baffled Reactor, reactive crystalization.

## 1. Introduction

Precipitation processes in electrolytic solutions are integral to various parts of the chemical industry, including water treatment and pharmaceutical production. These processes are characterized by dynamic changes in reactions and equilibria, necessitating sophisticated modeling to enhance production efficiency and quality. In these processes, changes in variables like compound concentration, temperature, or pressure shift electrolytic equilibria, leading to the precipitation of insoluble compounds due to a supersaturation of the solution. The dynamics of supersaturation play a crucial role in defining the nature and characteristics of the resulting crystals.

This work builds upon the dynamic modeling approach for systems with various equilibria, as initially established by Moe et al. (1995). The applicability of this methodology is demonstrated through its adoption by Kakhu et al. (2003), who incorporated phase appearance and disappearance using state-transition networks, and by Bremen et al. (2022), who integrated this concept into an open-source Modelica library. The approach presented here combines Moe et al.'s method with a population balance model.

This work's main objective is to develop a dynamic model that will be the basis for the optimization and model-based control for a large-scale production of an aromatic amine in a continuous oscillatory baffled reactor (COBR).

## 2. Model

A key complexity in modeling electrolyte systems lies in determining the complete set of concentrations of all species present in the system. The concentrations in these systems are primarily determined by equilibrium reactions, commonly assumed as rapid and reversible processes. In aqueous systems, the autoprotolysis of water must be accounted for, represented by the equilibrium: $2 \cdot H_2O \rightleftharpoons H_3O^+ + OH^-$. Commonly, for simplification, this reaction is exchanged by assuming the formation of free protons ($H^+$) instead of the hydronium ion. The dissociation of acids and bases then follows the equilibrium reactions: $HA \rightleftharpoons H^+ + A^-$; $OHB \rightleftharpoons OH^- + B^+$. The species involved in these equilibrium reactions conform to the equilibria according to the formula:

$$ln\left(K_j\right) = \sum_{i \in \mathcal{R}} \nu_{ij} \, ln(a_i),$$ (1)

where $K_j$ is the equilibrium constant for equilibrium $j$, with $\mathcal{R}$ representing the set of reactants, $a_i$ is the activity of species $i$, which is usually modelled using activity coefficient models, and $\nu_{ij}$ is the stoichiometric coefficient of species $i$ in reaction $j$.

The second key aspect in describing precipitation in electrolyte solutions involves modeling the particulate system of crystals and their characteristics. This is achieved using a population balance equation (PBE) model, which describes the evolution of the particle size distribution. The PBE is expressed as:

$$\frac{\partial f}{\partial t} + \frac{\partial Gf}{\partial L} = B.$$ (2)

In this equation, $f$ represents the number density function of the particles, $L$ is the characteristic length of the crystals, $G$ is the crystal growth rate, and $B$ is the nucleation rate for crystals of infinitesimal size. Employing a PBE model, the rate at which the crystallizing compound precipitates out of the solution can be determined as (Jha et al. (2017)):

$$r_c = 3 \frac{\rho_c k_v}{M_c} \int_0^\infty G \, f L^2 dL.$$ (3)

Here, $\rho_c$ denotes the density of the crystals, $k_v$ is the volumetric shape factor, and $M_c$ is the molar mass of the crystallizing compound. The combination of both the electrolyte and particulate systems is achieved by considering the component balances of the electrolytic species present as:

$$\frac{d\bar{n}}{dt} = \dot{\bar{n}}_{in} - \dot{\bar{n}}_{out} + V\left(\bar{\nu}_{eq}^T \bar{r}_{eq} + \bar{\nu}_r^T \bar{r}_r\right).$$ (4)

In this balance, $\bar{n}$ is the vector of moles of components in the systems and $\dot{\bar{n}}_{in}$ and $\dot{\bar{n}}_{out}$ represent the rates of flow entering and exiting the system, respectively. The second term on the right-hand side accounts for the equilibrium reactions (denoted by the subscript $eq$) and rate terms for slow processes (denoted by the subscript $r$), such as crystallization, where $\bar{\nu}$ are matricies of stoiciometric coefficients and $\bar{r}$ are the vectors of volume specific reactions rates. This balance is lumped over the entire reaction volume $V$.

*2.1. Index reduction*

Addressing the high index of the set of differential algebraic equations derived from the equilibrium equations Eq. 1 and the component balance equations Eq. 4, an index reduction strategy is employed to formulate a semi-explicit system with index one. This approach introduces reaction invariants related to the equilibrium reactions, initially presented by Asbjørnsen and Field 1970 and later applied by Moe et al. 1995. In the first step, it involves determining the basis vectors for the nullspace, $\lambda_k$, corresponding to the stoichiometric coefficient matrix associated with these equilibrium reactions:

$$\bar{v}_{eq} \lambda_k = 0 \quad for \ k = 1, \dots, N_{comp} - N_{eq}, \tag{5}$$

where $N_{comp}$ represents the total number of species in the equilibria, and $N_{eq}$ is the count of equilibria considered. The equilibrium reaction invariant states, denoted as $\tilde{n}$, are subsequently defined through:

$$\tilde{n}_k = \lambda_k{}^T \tilde{n} \quad for \ k = 1, \dots, N_{comp} - N_{eq}. \tag{6}$$

The dynamics governing these invariant states are captured by differential equations:

$$\frac{d\tilde{n}_k}{dt} = \lambda_k{}^T (\dot{n}_{in} - \dot{n}_{out} + \bar{v}_r^T \bar{r}_r V) \quad for \ k = 1, \dots, N_{comp} - N_{eq}. \tag{7}$$

By maintaining Eq. 1, alongside Eq. 6, and Eq. 7 where $\tilde{n}$ are considered as algebraic variables, the desired index reduction is achieved, while maintaining the component balances and equilibria and disregarding the dynamics of the equilibrium reactions.

## 2.2. Method of moments

Considering the partial differential PBE to describe the particulate system often becomes computationally intractable. Even in its discretized form, the complete model is likely not feasible for real-time application, particularly in a distributed reaction system like a COBR. This is the reason for employing the method of moments, which introduces moments of order $p$, $\mu_p$, as differential states, defined as:

$$\mu_p = \int_0^\infty L^p \, f \, dL. \tag{8}$$

Assuming that the growth rate, $G$, is independent of particle size, and that there is no particle agglomeration or breakage, the resulting moment balances are:

$$\frac{d\mu_0}{dt} = B; \ \frac{d\mu_p}{dt} = pG\mu_{p-1} \quad for \ p \in \mathbb{Z}^+ \backslash \{0\}. \tag{9}$$

To model the dependence of the growth rate $G$ and nucleation rate $B$ on supersaturation, the following exponential relationships are introduced:

$$G = k_g S^{\sigma_g}; \ B = k_b S^{\sigma_b}. \tag{10}$$

The supersaturation, $S$, is defined as:

$$S = max\left(\frac{C - C_{sol}}{C_{sol}}, 0\right). \tag{11}$$

Here, $C$ is the concentration of the precipitating solute, and $C_{sol}$ is its concentration at equilibrium, which is a function of the system state. While the use of the maximum-function in the definition of supersaturation is important for parameter estimation to avoid negative values, it introduces non-smoothness in the solution of the model. To address this, the function is approximated as follows:

$$max(x,y) \approx \frac{x+y+\sqrt{(x-y)^2+2k^2}}{2}. \tag{12}$$

In this approximation, $k$ acts as a smoothing parameter, such that as $k \to 0$, the maximum-function is accurately approximated.

### 2.3. Continuous Oscillatory Baffled Reactor Model

In modeling precipitation within a COBR, a plug flow reactor model is assumed, based on the COBR's effective mixing characteristics (Stonestreet and Van Der Veeken, 1999), which lead to the assumption of negligible backmixing. The model is discretized using a backward difference scheme, effectively transforming it into a sequence of small; in-series connected continuously stirred tank reactors. Side-injections are modelled as feed streams into single reactor elements. A schematic view of the modelled COBR is depicted in Figure 1.

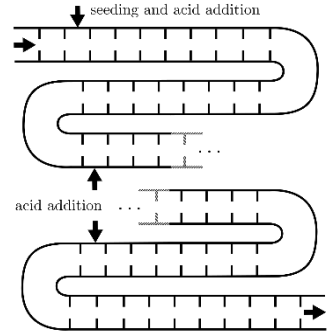

Figure 1: COBR with acid side injections.

## 3. Case Study

The case study explores the industrial crystallization of an aromatic amine. Initially, the amine is dissolved in a basic solution. The addition of acid neutralizes this solution, leading to the protonation of the amine, which then precipitates due to its low solubility in its protonated form. The primary objective is to enhance manufacturing efficiency by shifting from batch processing to continuous production methods. This change aims to achieve greater space efficiency, environmental sustainability, and cost reduction. A crucial part of this study is the optimization of the crystallization process. Key factors influencing this include the acid addition, which causes solution supersaturation, and the control of process variables like temperature.

For simplification, it is assumed that strong acids and bases fully dissociate and salts completely dissolve in the solution, predominantly forming ionic species. Thus, adding hydrochloric acid is treated as directly producing hydronium and chloride ions, enabling the disregard of equilibrium equations for these reactions. Additionally, due to current unavailability of thermodynamic data, the activities of the components are assumed to be represented by their concentrations. The equilibrium constant of the acidic and basic components of the system are derived from known pKa and pKb values.

The model is implemented in the open-source Python library *Pyomo*, utilizing its modules, *pyomo.dae* (Nicholson et. al, 2018) and *pyomo.paramest* (Klise et. al, 2019).

Parameter estimation for the model is conducted using data from isothermal semi-batch experiments where hydrochloric acid is continuously added to the process solution. A solubility function specific to the neutralized form of the amine is used to calculate the supersaturation. The model is discretized using orthogonal collocation on finite elements. The parameter estimation problem is structured using a least squares objective function, aiming to minimize the difference between the model's predicted and the experimentally measured time-dependent concentration of the amine in the solution. The parameters to be determined include the equilibrium constant for the amine's protonation and the parameters that describe the crystallization kinetics, as indicated in Eq. 10. Given the highly non-linear nature of the optimization problem and the lack of a solver with global guarantees, the parameters are initially set to random values within a predefined range, and the optimization process is repeated 5000 times to ensure a good fit.

## 4. Results

### *4.1. Parameter Estimation*

The parameter estimation process results in a model that closely matches the experimental data, with an average relative error of 8%. Figure 2 shows the normalized results for one of the ten experiments conducted. The lower section of the figure illustrates the concentration of free protons, highlighting the pH change during the acidification and precipitation.

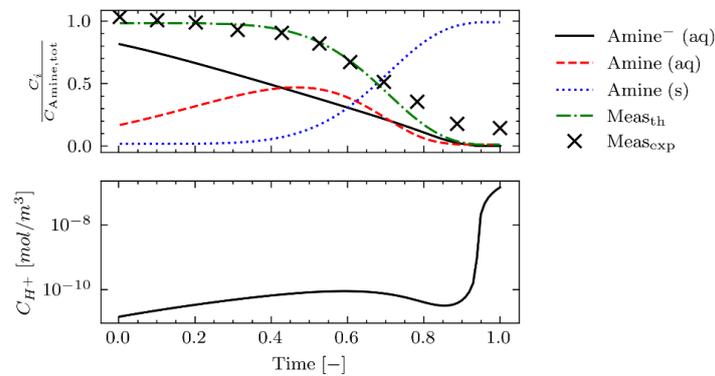

Figure 2: Relative Concentration profiles of the aromatic amine in the process solution (top) and the free protons (bottom).

While the model effectively captures the overall trend of the measurements, it deviates towards the end of the trajectory, predicting lower amine concentrations than observed. This discrepancy could be attributed to a potential inaccuracy in the solubility function, which may estimate a lower solubility than what actually occurs in the experiments.

### *4.2. Continuous Oscillatory Baffled Reactor Simulation*

In the COBR simulation, six hydrochloric acid addition flows are chosen to maintain consistent supersaturation and ensure complete conversion of the aromatic amine. The simulation results are illustrated in Figure 3. This figure is generated from a simulation where the COBR operates until it reaches a steady state, where most of the aromatic amine is precipitated from the solution.

The profile for adding hydrochloric acid is designed to initially introduce approximately 40% of the total acid flow, to achieve a desired level of supersaturation, followed by a period of reduced addition, to maintain the desired level. Towards the end of the process, the acid feed rate is significantly increased. This change corresponds to the increasing rate of precipitation, which is influenced by the growing surface area of the forming and expanding crystals. The crystallization process inherently leads to sharp peaks in supersaturation. These peaks can cause scaling on certain sections of the reactor walls, a challenge that has also been observed in laboratory experiments. This aspect is a critical consideration in the design of the acid feed profile for the reactor.

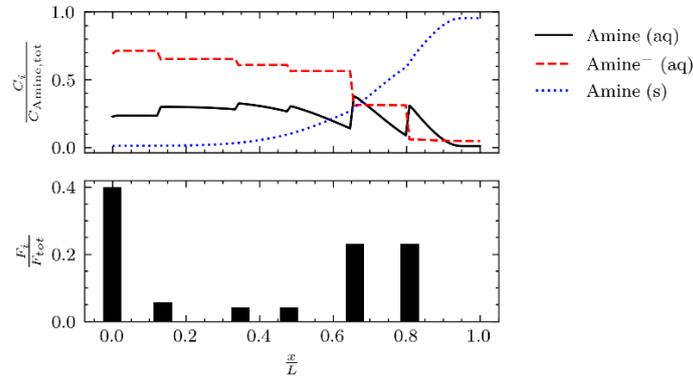

Figure 3: Relative concentration profiles of the aromatic amine along the reactor of length L at steady state (top) and the hydrochloric acid side feed flows (bottom).

## 5. Conclusion

This study successfully developed a dynamic model for the precipitation in electrolyte systems, which is applied in a batch reactor and COBR. The model, which effectively integrates equilibrium reactions and crystallization kinetics, demonstrates strong agreement with experimental data. The application of the model in the COBR simulation highlights its potential for optimizing and controlling a continuous production process. The insights gained, particularly in controlling the acid addition and addressing scaling issues, provide valuable guidance for enhancing manufacturing efficiency and environmental sustainability in large-scale industrial applications.

## Acknowledgement

The project leading to this publication has received funding from the European Union's Horizon research and innovation programme under grant agreement No 101058279.

## References

Asbjørnsen O. A., Field M., 1970, Response modes of continuous stirred tank reactors. Chemical Engineering Science, *25*(11), 1627-1636.

Bremen, A.M., Ebeling, K.M., Schulte, V., Pavšek, J., Mitsos, A., 2022, Dynamic modeling of aqueous electrolyte systems in Modelica, Computers & Chemical Engineering 166.

Jha S.K., Karthika S., Radhakrishnan T.K., 2017. Modelling and control of crystallization process, Resource-Efficient Technologies 3, 94–100.

Klise K.A., Nicholson B.L., Staid A., Woodruff D.L., 2019, Parmest: Parameter Estimation Via Pyomo, Computer Aided Chemical Engineering. Elsevier, pp. 41–46.

Kakhu A.I., Pantelides, C.C., 2003. Dynamic modelling of aqueous electrolyte systems, Computers & Chemical Engineering 27, 869–882.

Moe H.I., Hauan S., Lien K.M., Hertzberg T., 1995, Dynamic model of a system with phase- and reaction equilibrium, European Symposium on Computer Aided Process Engineering.

Nicholson B., Siirola J.D., Watson J.-P., Zavala V.M., Biegler L.T., 2018, pyomo.dae: a modeling and automatic discretization framework for optimization with differential and algebraic equations, Math. Prog. Comp. 10, 187–223.

Ramkrishna D., 2000, Population balances: theory and applications to particulate systems in engineering, Academic Press, San Diego, CA.

*Dynamic Modeling of Precipitation in Electrolyte Systems*

Stonestreet P., Van Der Veeken P.M.J., 1999, The Effects of Oscillatory Flow and Bulk Flow
    Components on Residence Time Distribution in Baffled Tube Reactors, Chemical Engineering
    Research and Design 77, 671–684.